\documentclass[fleqn,usenatbib,useAMS]{mnras}

\usepackage{graphicx}	% Including figure files
\usepackage{amsmath}	% Advanced maths commands
\usepackage{amssymb}	% Extra maths symbols
\usepackage{multicol}        % Multi-column entries in tables
\usepackage{bm}		% Bold maths symbols, including upright Greek
\usepackage{pdflscape}	% Landscape pages
\usepackage{csvsimple}

\usepackage[T1]{fontenc}
\usepackage{ae,aecompl}

\usepackage{newtxtext,newtxmath}
\title[PSR~J0908--4913 and its SNR]
{A supernova remnant association for the fast-moving pulsar PSR~J0908--4913}
\author[S.~Johnston and M.~E. Lower]
{Simon Johnston$^{1}$\thanks{E-mail: simon.johnston@csiro.au}
and Marcus~E. Lower$^{2,1}$
\\
$^{1}$Australia Telescope National Facility, CSIRO Space and Astronomy, PO~Box~76, Epping NSW~1710, Australia\\
$^{2}$Centre for Astrophysics and Supercomputing, Swinburne University of Technology, PO Box 218, Hawthorn, VIC 3122, Australia\\
}

\date{Last updated; in original form}

\pubyear{2021}

\begin{document}
\label{firstpage}
\pagerange{\pageref{firstpage}--\pageref{lastpage}}
\maketitle

\begin{abstract}
A recent measurement of the proper motion of PSR~J0908--4913 shows that it is a fast moving object at a distance of some 3~kpc. Here we present evidence that the pulsar is located at the edge of a previously unknown, filled-centre supernova remnant, G270.4--1.0. The velocity vector of the pulsar points directly away from the centre of the remnant. The putative association of the pulsar with SNR G270.4--1.0 implies the pulsar is $\sim$12~kyr old, significantly less than its characteristic age of 110~kyr. We show that the rotation axis of the pulsar points in the direction of the proper motion. Rotation measure and dispersion measure variations are seen over time, likely indicating the pulsar is passing behind a filament of the remnant.
\end{abstract}

\begin{keywords}
pulsars:general -- pulsars: individual: PSR~J0908--4913
\end{keywords}

%%%%%%%%%%%%%%%%%%%%%%%%%%%%%%%%%%%%%%%%%%%%%%%%%%

%%%%%%%%%%%%%%%%% BODY OF PAPER %%%%%%%%%%%%%%%%%%
\section{Introduction}
PSR~J0908--4913 was first discovered by \citet{dmd+88}. It has a spin period of 106~ms and a dispersion measure (DM) of 180.4~cm$^{-3}$pc. The spin-down energy of the pulsar is high ($4.9\times10^{35}$~ergs$^{-1}$) and its characteristic age is 110~kyr. As pulsars go, this particular object has a lot to offer. First, it is one of the few pulsars for which we see radio emission from both magnetic poles (as originally identified by \citealt{dmd+88}). This means that the geometry of the pulsar is well determined \citep{kj08,jk19}, another rarity in the pulsar population. It emits gamma-rays \citep{2pc}, as expected for highly-energetic pulsars which are orthogonal rotators \citep{jskk20}, and has been weakly detected at X-ray wavelengths \citep{kdpg12}. The pulsar powers a pulsar wind nebula and its motion through the interstellar medium produces a radio bow-shock \citep{gsfj98}. Finally, large DM variations are seen \citep{pkj+13}, making the pulsar an excellent probe of the interstellar medium. Most recently, measurements of its proper motion imply it is a fast-moving object \citet{ljd+21}, contrary to previous results \citep{jnk98}.

In this paper we explore the implications of the newly measured proper motion, describe a putative supernova remnant association and present the dispersion measure and rotation measure history of the pulsar over 15 years' worth of observations.

\section{Distance}
The distance to PSR~J0908--4913 has not been directly determined through parallax measurements. We are therefore reliant on distance estimates based either on the pulsar's dispersion measure or spectroscopy and, for this pulsar, these distance estimates have a chequered history. In the discovery paper, \citet{dmd+88} estimated a distance of 3~kpc based on the DM. The next iteration of distance models increased this estimate to 6.6~kpc \citep{tc93}. H{\sc i} measurements in the direction of the pulsar by \citet{kjww95} showed clear absorption features which enabled them to place a lower distance limit of 2.4~kpc and an upper distance limit of 6.7~kpc to the pulsar. \citet{cl02} incorporated the H{\sc i} limits into their model and their estimate to the distance of the pulsar was only 2.6~kpc. Finally, in the latest published model of the Galactic electron density, \citet{ymw17} have a distance estimate of a mere 1.0~kpc, clearly not compatible with the H{\sc i} lower limit.

One final piece of information is the detection of pulsed gamma-ray emission from the pulsar \citep{2pc}. At a distance of 1.0~kpc, the efficiency of the pulsar in gamma-rays is only 0.6\%, at the low end of the scale for pulsars of this type \citep{2pc}. A distance of 2.5 to 3.0~kpc raises the efficiency to 4-5\%, more in line with expectations. In X-rays this works in the opposite direction. \citet{kdpg12} report an X-ray efficiency of $5.7\times10^{-5}$ at a distance of 6.6~kpc, and this decreases to only $1.2\times10^{-5}$ with the closer distance estimate. We return to the distance problem in the next section.

\section{Space Velocity}
A comprehensive study of the scintillation of southern pulsars, including PSR~J0908--4913, was carried out by \citet{jnk98}. From a single 3~h observation of the pulsar at 1520~MHz, they derived a scintillation bandwidth of 12~MHz and a very long scintillation timescale of 4770~s. This implies a scintillation velocity of only $\sim$50~kms$^{-1}$ at their adopted distance of 6.7~kpc (and hence an even lower value if the distance is 3~kpc).

It became apparent that the scintillation bandwidth as derived by \citet{jnk98} was in error. In particular, evidence for a scatter-broadened profile at lower frequencies indicates a much smaller value for the scintillation bandwidth than they derived. In the geometric modelling by \citet{kj08}, they estimated a scintillation bandwidth of only 2500~Hz at 1.4~GHz. Additionally, there are no obvious scintillation bands present in recent data taken with the Parkes telescope at frequencies between 700 and 4000~MHz \citep{jsd+21}, while data taken at 8.5~GHz \citep{jkw06} showed narrow scintillation structures. To clear up the scintillation question, we collected data using the Parkes radio telescope at an observing frequency of 8.5~GHz on 2021 Feb 26. The total bandwidth was 1024~MHz subdivided into 1024 channels each of width 1~MHz, with a total integration time of 6930~s subdivided into 30~s blocks. Polarization calibration was performed via observations of a pulsed square-wave injected directly into the feed. Flux calibration was carried out using observations of PKS~B1934--638 which has a flux density of 2.8~Jy at 8.5~GHz. The flux density of the pulsar was computed for each time-step in each frequency channel to create a dynamic spectrum. These data were then analysed using {\sc scintools}\footnote{https://github.com/danielreardon/scintools} as described in \citet{rcb+20}. The decorrelation bandwidth, $\Delta\nu_{d}$, was computed to be 3.2~MHz and the scintillation time, $t_{d}$, to be 59~s. As a check of these numbers, we obtained the 2005 data at 8.356~GHz from the Parkes pulsar data archive \footnote{https://data.csiro.au/dap/} and applied the same software tools. This yielded 2.8~MHz and 56~s for the scintillation bandwidth and time, both in good agreement with the more recent data. The scintillation velocity, $V_{\rm ISS}$ is given by
\begin{equation}
V_{\rm ISS} = A_{V} \; \frac{(D\Delta\nu_{d})^{1/2}}{\nu\, t_{d}},
\end{equation}
where $D$ is the distance to the pulsar in kpc, $\nu$ is the observing frequency in GHz, $\Delta\nu_{d}$ is in MHz and $t_{d}$ is in s. The constant $A_{V}$ is derived by assuming a Kolmogorov turbulence spectrum and a homogeneously turbulent medium, and has been estimated as $3.85\times 10^{4}$. For $D=3$~kpc, using the values determined above yields $V_{\rm ISS} = 250$~kms$^{-1}$.

Modern inference techniques have been brought to bear on the timing of the pulsar over two decades. This enabled \citet{ljd+21} to determine a proper motion of $-37\pm9$~mas yr$^{-1}$ in right ascension and $31\pm10$~mas yr$^{-1}$ in declination. This leads to an overall proper motion of $47\pm9$~mas yr$^{-1}$, hence a transverse velocity of $670\pm130$~kms$^{-1}$ assuming a distance of 3~kpc at an angle of $50\degr \pm 11\degr$ measured from north towards west. We note that the velocity is already high compared to the bulk of the pulsar population (e.g. \citealt{hllk05}) and it therefore seems unlikely that the pulsar can be as distant as 7~kpc. In Galactic coordinates, the pulsar's velocity is oriented almost entirely in the negative longitude direction with only a small component in the latitude direction. With the pulsar now more than 1\degr\ (60~pc at a distance of 3~kpc) from the Galactic plane, it seems as if it did not originate from zero latitude. It should be noted that the bulk of the emission from the Galactic plane at these longitudes is at negative latitudes.

There remains a significant discrepancy between the scintillation velocity and the proper motion. One way to reconcile the two values would be to postulate that the scattering screen lies significantly closer to the pulsar than to the observer.

\section{Wind Nebula}
\begin{figure}
\begin{center}
\begin{tabular}{c}
\includegraphics[width=8.2cm,angle=0]{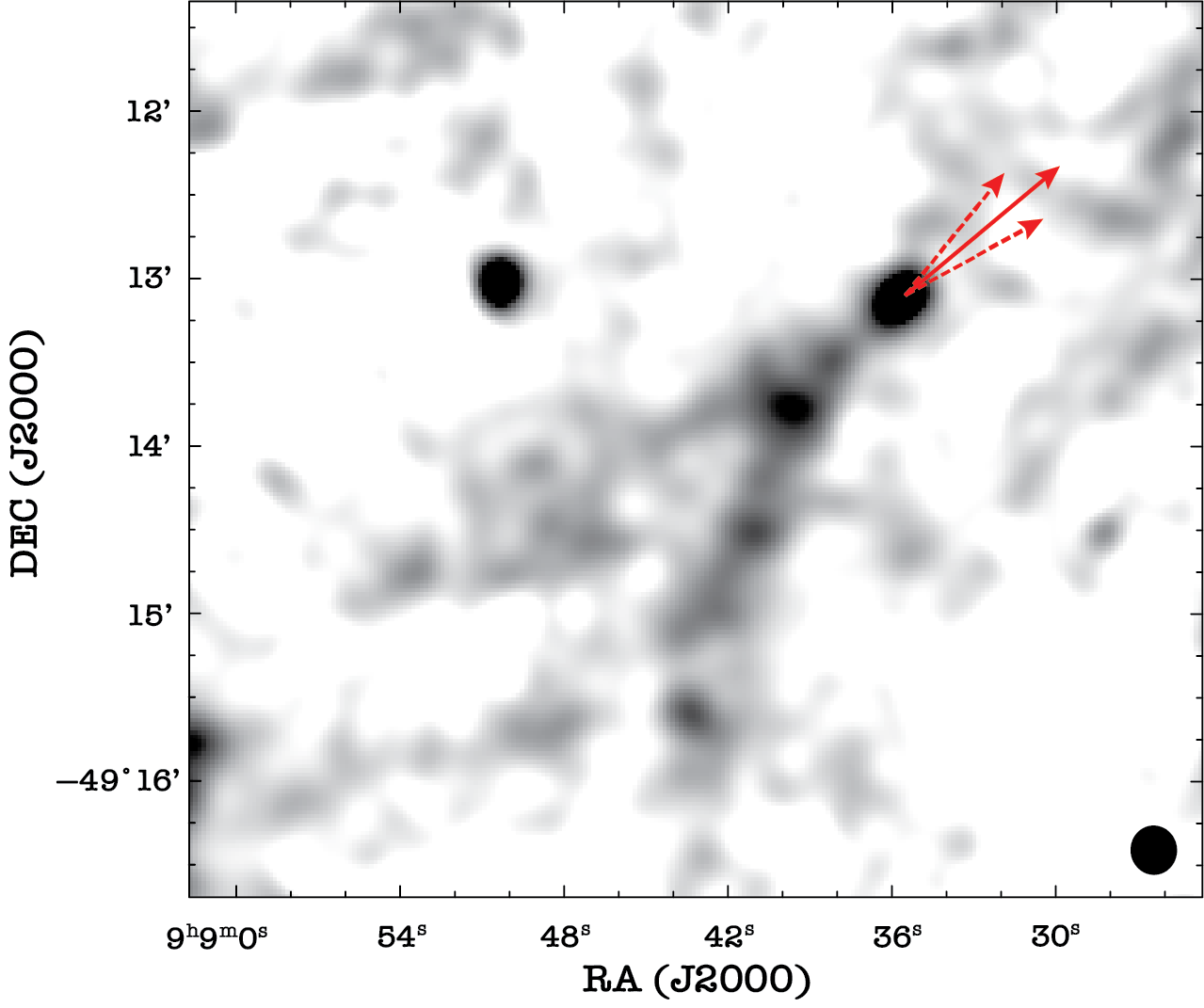} \\
\end{tabular}
\end{center}
\caption{Radio image at 1.3\,GHz from observations in 1997 with the Australia Telescope Compact Array, adapted from \citet{gsfj98}. The synthesised beam ($17.7'' \times 16.6''$ at position angle 5\degr) is shown in the bottom right. Proper-motion direction from \citet{ljd+21} is shown by the solid-red arrow. Uncertainty in the proper motion is represented by the dashed arrows.}
\label{fig:0908image}
\end{figure}
\begin{figure}
\begin{center}
\begin{tabular}{c}
\includegraphics[width=8.2cm,angle=0]{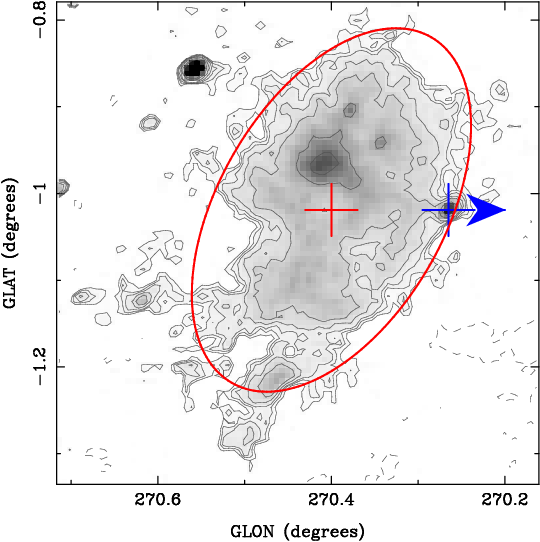} \\
\end{tabular}
\end{center}
\caption{Radio image at 843~MHz in Galactic coordinates taken from the Molonglo Galactic Plane Survey of the region surrounding PSR~J0908--4913. Observations were taken in 1991, the synthesised beam measures $43'' \times 57''$.
The blue cross denotes the location of the pulsar with the arrow showing the direction of the proper motion. The red ellipse marks the approximate region of the SNR, with the centre of the ellipse shown by the red cross.
%The grey-scale bar shows the flux density in Jy/beam.
}
\label{fig:0908mgps}
\end{figure}
More than twenty years ago, a search for pulsar wind nebulae (PWN) around southern \citep{sgj99} and northern pulsars \citep{fs97} was carried out. Although some 60 sources were surveyed, only one clear example of a pulsar harbouring a PWN, PSR~J0908--4913 \citep{gsfj98}, was found. In addition to the PWN surrounding the pulsar, a clear trail or bow-shock was seen in the radio image which is reproduced in Figure~\ref{fig:0908image}. The pulsar is one of only a small handful of pulsars known to have radio PWN. The \citet{gsfj98} interpretation of these structures was based on a distance estimate of 7~kpc and a low velocity of 60~kms$^{-1}$. This led to the conclusion that the pulsar was moving through a very dense medium (hydrogen density $n_{\rm H}$ of 2~cm$^{-3}$) and that the bow-shock emission was the same age as the pulsar ($\sim$10$^5$~yr). Figure~\ref{fig:0908image} also shows the proper motion vector as determined by \citet{ljd+21}, pointing in exactly the direction predicted by \citet{gsfj98}.

Our understanding now is that the pulsar is moving some ten times faster than they assumed and that the distance is less than half the previous value. This leads to a very different interpretation. The PWN is unresolved at this resolution, so we assume the stand-off distance, $r_s$ must be $< 1.2\times10^{15}~d$~cm (with $d$ in kpc). If we assume that the entire spin-down energy of the pulsar goes into the PWN and we equate this with the ram pressure due to the pulsar's velocity, then we can compute the hydrogen density must be $n_{\rm H} \sim 0.08$~cm$^{-3}$. This value can be higher if $r_s$ is smaller than assumed but can also be lower if the efficiency of the pulsar in producing a particle wind is not 100\%. With those caveats, we no longer need the extreme value of $n_{\rm H}$ derived by \citet{gsfj98} and our value is more in line with that expected from the interstellar medium in general (e.g. \citealt{mo77}). Furthermore, we see that the bow-shock length of 3 arcmin equates to some 2.5~pc assuming a distance of 3~kpc. With a pulsar velocity of 670~kms$^{-1}$, the duration of the emission from the bow-shock must be less than $10^4$~yr before it fades into the medium. Two explanations for this are possible. First, the observations may not be sensitive enough to show the low surface-brightness structures far from the pulsar. Secondly, we do not have a good handle on the spectrum of the PWN in order to determine a break frequency from the ageing of the synchrotron electrons. This could in turn allow us to determine the magnetic field within the PWN: a high value (above 1~mG) would strongly limit the lifetime. We note that any X-ray nebula must be faint as \citet{kdpg12} report only 10 photons from a 35~ks \textit{Chandra} observation of the region around the pulsar.

However, it appears to be the case that the structure seen in the high resolution image is only part of a much larger, low surface brightness structure. A radio image from the Molonglo Galactic Plane Survey \citep{gcly99,mmg+07} at the pulsar's location is shown in Figure~\ref{fig:0908mgps}. Could this be a filled-centre supernova remnant (SNR) created by the same supernova that formed the pulsar? The pulsar is currently situated on the edge of the structure with its proper motion vector points directly away from the centre of emission, lending credence to a link between the two. We also note the similarities between this and several other systems: SNR~W44 with PSR~B1853+01 \citep{jsa93,fggd96}, CTB~80 with PSR~B1951+32 \citep{cdg+03}, PSR~J0002+6216 with CTB~1 \citep{skr+19}, PSR~J0538+2817 with S147 \citep{klh+03,nrb+07} and SNR~G5.4--1.2 with PSR~B1757--24 \citep{mkj+91,gf00}. In W44, CTB~80 and S147 the pulsar is embedded within the shell, whereas in G5.4--1.2, the pulsar has just exited its parent shell and in CTB~1 the high-velocity pulsar has escaped its shell entirely. In all five of these systems, an elongated PWN along the direction of motion is seen.

We therefore assume that the structure seen in Figure~\ref{fig:0908mgps} is indeed a supernova remnant, to which we assign the name SNR~G320.4--1.0. The SNR is roughly ellipsoidal, and an estimate of its centre is shown by the red cross in Figure~\ref{fig:0908mgps}. If the pulsar was born in the same explosion as created the SNR, its current displacement of some 10 arcmin from the centre of the SNR and its proper motion of $\sim$47~masyr$^{-1}$ yield a kinematic age of $\sim$12.7~kyr (a value which is independent of the distance). The characteristic age, $\tau_c$, of the pulsar given by $\tau_c = P/(2\dot{P})$ where $P$ is the spin period and $\dot{P}$ its derivative is 110~kyr, considerably higher than the kinematic age. In the examples of similar systems mentioned above, that the kinematic age of PSR~B1757--24 is much larger than its $\tau_c$ \citep{gf00}, with the opposite being true for PSR~B1951+32 \citep{zbcg08}, PSR~J0002+6216 \citep{skr+19} and PSR~J0538+2817 \citep{klh+03}, while for PSR~B1853+01 the ages roughly agree. It should be noted that $\tau_c$ comes with several caveats; it assumes a very fast initial spin period and a braking index of $n=3$, neither of which are well established \citep{jk17,pjs+20}. To reconcile $\tau_c$ with the kinematic age in PSR~J0908--4913 then either it was born spinning with a period not too different from its current spin period and has $n=3$, or it was born spinning fast and has $n$ of order 20 \citep{jk17}. Curiously and perhaps coincidentally, a value of $n=23$ has been determined by \citet{ljd+21} following nearly three decades of timing this pulsar.
\begin{figure}
\begin{center}
\begin{tabular}{c}
\includegraphics[width=8.2cm,angle=0]{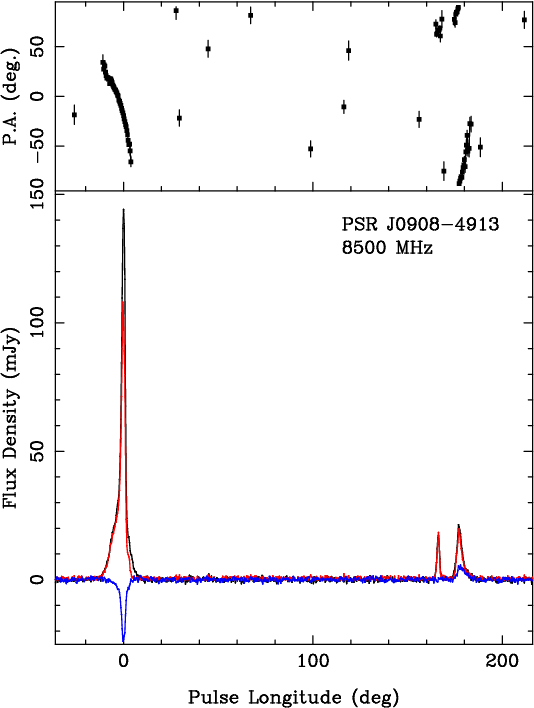} \\
\end{tabular}
\end{center}
\caption{Flux and polarization calibrated profile of PSR~J0908--4913 at 8.5~GHz. The bottom panel shows total intensity (black), linear polarization (red) and circular polarization (blue). The top panel shows the position angle of the linear polarization at infinite frequency.}
\label{fig:0908poln}
\end{figure}

\section{Velocity vector and the orientation of the rotation axis}
There exists a remarkable correlation between the orientation of a pulsar's rotation axis and its direction of proper motion. First found in radio pulsars by \citet{jhv+05} based on a theoretical prediction by \citet{sp98} and later confirmed via other radio \citep{ran07,nsk+13} and X-ray \citep{hgh01,nr04} datasets, it appears to hold true for most, if not all, pulsars with an age less than about 10$^{5}$~yr. The geometry of PSR~J0908--4913 is well established \citep{kj08,jk19}. In brief, the angle between the rotation axis and the magnetic axis is 83.59\degr\ and that between the rotation axis and the line of sight is 91.03\degr. Figure~\ref{fig:0908poln} shows the polarization profile of the pulsar at 8.5~GHz. At the location of the magnetic pole identified by \citet{jk19}, the position angle of the linear polarization is --38\degr. Remarkably, this is 88\degr\ away from the position angle of the proper motion. PSR~J0908--4913 therefore emits in a mode orthogonal to the magnetic field lines and shows that the rotation axis and the proper motion vector are aligned (c.f. the Vela pulsar, \citealt{jhv+05}).

\section{DM and RM variability}
\begin{figure}
\begin{center}
\begin{tabular}{c}
\includegraphics[width=8.2cm,angle=0]{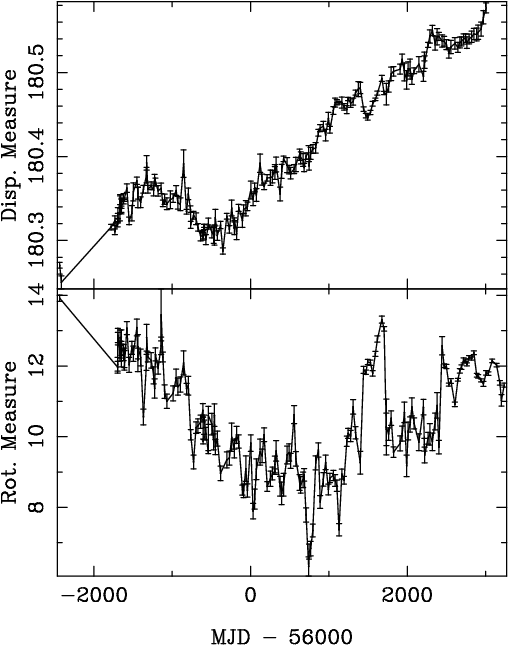} \\
\end{tabular}
\end{center}
\caption{Dispersion Measure (top panel) and Rotation Measure (bottom panel) as a function of MJD for PSR~J0908--4913.}
\label{fig:0908dmrm}
\end{figure}
Given the high proper motion of the pulsar, and its relatively close distance, we might expect to see variations in the pulsar's dispersion measure and rotation measure (RM) over time. Indeed, in a comprehensive survey for DM variations carried out by \citet{pkj+13}, they found significant variations in the DM towards this pulsar. After a gap of nearly 8 years, \citet{jsd+21} showed that the DM variations had changed sign since the previous measurements.

The top panel in Figure~\ref{fig:0908dmrm} shows the entire DM history (186 independent measurements) of this pulsar from MJD 53564 (2005 July) to MJD 59237 (2021 January).  Apart from an initial gap at the beginning of the series, data have been taken with a monthly cadence since 2007 April. \citet{pkj+13} used data between MJD 54220 and 56200 where the DM was largely decreasing. Since MJD 56000, the DM has been steadily increasing at a mean rate of 0.029~cm$^{-3}$pcyr$^{-1}$. %although there is a hint that it has begun to drop again in the last 200~days.
The RM towards the pulsar is rather low, yielding a mean magnetic field strength of $<0.1~\mu$G along the line of sight. This indicates the likely presence of reversals in the magnetic field direction in this direction. The RM history of the pulsar is given in the bottom panel of Figure~\ref{fig:0908dmrm}. Significant RM variations are seen, first a drop from 13~radm$^{-2}$ to 9~radm$^{-2}$ followed by a rise back to 12~radm$^{-2}$.

Both the DM and RM variations could be caused by the pulsar motion behind a filament from the SNR. Over the course of the 10~yr monitoring program, the pulsar has traversed a distance of some 7~mpc and in this time the DM has changed by 0.25~cm$^{-3}$pc. This yields an electron density excess of some 35~cm$^{-3}$. With an RM change of 4~radm$^{-2}$ over the same time period, this would imply a filament magnetic field of 20~$\mu$G. These values are similar to those derived following RM and DM changes in the Vela pulsar \citep{hhc85} and in-line with what is expected for SNRs generally \citep{rgb12}.

\section{Summary and next steps}
More than 30 years after its original discovery, a complete picture of the remarkable pulsar PSR~J0908--4913 is now beginning to emerge. Recent measurements of the pulsar's proper motion show it to be a high velocity object, and the direction of its motion is exactly as predicted from the morphology of its associated bow-shock nebula \citep{gsfj98}. The orientation of the rotation axis is aligned with the proper motion vector once orthogonal-mode emission is taken into account. Examination of radio images of the Galactic plane in the vicinity of the pulsar reveal the presence of a putative SNR, G320.4--1.0. The pulsar appears to be on the edge of the SNR shell, with the proper motion vector pointing directly away from the SNR centre. If the association between G320.4--1.0 and pulsar is correct, then the pulsar can only be $\sim$12~kyr old, significantly less than its characteristic age. Our preferred distance to the pulsar is 3~kpc, based on the lower limit from H{\sc i} absorption and to ensure a velocity of less than 1000~kms$^{-1}$.

We plan to carry out very long baseline interferometry (VLBI) to obtain high angular resolution images of the PWN. This should allow us to determine the bow-shock stand-off radius in order to better understand the energetics of the system. X-ray observations in this direction are also warranted to supplement the previous data from \citet{kdpg12}. Deep X-ray observations may well allow better constraints on the PWN and the relationship with the SNR, as has been done recently for PSR~J1709--4429 and its putative parent remnant SNR~G343.1--2.3 \citep{vrk+21}.

Finally we note that \citet{yzm+21} have recently determined a full three dimensional picture of the motion of PSR~J0538+2817 within its SNR via a combination of radio polarization and scintillation measurements. Such an approach should also be possible for PSR~J0908--4913 through sensitive observations using the pulsar program on MeerKAT \citep{jkk+20}.

\section{Data availability}
Pulsar data taken with the Parkes telescope are available via a public archive\footnote{https://data.csiro.au/dap/}. Data taken with the Compact Array are available via a public archive\footnote{https://atoa.atnf.csiro.au/}. The Molonglo Galactic Plane Survey is available via FITS files\footnote{http://www.astrop.physics.usyd.edu.au/MGPS/}.

\section*{Acknowledgements}
The Parkes telescope (\textit{Murriyang}) and the Australia Telescope Compact Array are part of the Australia Telescope National Facility which is funded by the Commonwealth of Australia for operation as a National Facility managed by CSIRO. We acknowledge the Wiradjuri and Gomeroi people as the traditional owners of the respective observatory sites.
M.E.L. receives support from Australian Research Council Laureate Fellowship FL150100148, the Australian Government Research Training Program and CSIRO Space and Astronomy.
We thank S. Mader for assistance with the 8~GHz observing, B. Posselt for useful comments and the referee for a timely review.

%%%%%%%%%%%%%%%%%%%%%%%%%%%%%%%%%%%%%%%%%%%%%%%%%%



\bibliographystyle{mnras}
\bibliography{0908} % if your bibtex file is called example.bib

%%%%%%%%%%%%%%%%%%%%%%%%%%%%%%%%%%%%%%%%%%%%%%%%%%

% Don't change these lines
\bsp	% typesetting comment
\label{lastpage}
\end{document}